\documentclass[aps,pra,twocolumn,amsfonts,amssymb,amsmath,showpacs,
floatfix,nofootinbib,groupedaddress,superscriptaddress,citesort]{revtex4}
\usepackage{mathrsfs}
\usepackage{amsfonts}
\usepackage{amstext}
\usepackage{amsmath}
\usepackage{amssymb}
\usepackage{bm}
\usepackage{bbm}
\usepackage[dvips]{graphicx}
\def\qed{\leavevmode\unskip\penalty9999 \hbox{}\nobreak\hfill
     \quad\hbox{\leavevmode  \hbox to.77778em{%
              \hfil\vrule   \vbox to.675em%
               {\hrule width.6em\vfil\hrule}\vrule\hfil}}
     \par\vskip3pt}

\begin{document}

\title{ Strictly incoherent operations for one-qubit systems}

\author{Shuanping Du}
\email{shuanpingdu@yahoo.com} \affiliation{School of Mathematical
Sciences, Xiamen University, Xiamen, Fujian, 361000, China}

\author{Zhaofang Bai}\thanks{Corresponding author}
\email{baizhaofang@xmu.edu.cn} \affiliation{School of Mathematical
Sciences, Xiamen University, Xiamen, Fujian, 361000, China}

\begin{abstract}

Strictly incoherent operations (SIO) proposed in [Phys. Rev. Lett. 116, 120404 (2016)] are promising to be a good
candidate of free operations in the resource theory of quantum coherence, setting against the central role of local operations and classical communication in the resource theory of quantum entanglement. An important open problem is an efficient description for strictly incoherent operations in physical region. Such a description plays key role for axiomatic study of resource theory of quantum coherence. We are aimed to give a structural characterization of bistochastic SIOs in terms of Pauli operators and the Phase operator for one-qubit systems.  Some applications of our results are also sketched in reconstructing quantum thermal averages via a quantum computer and in coherence manipulation.

\end{abstract}

\maketitle

{\it Introduction and main results.---} Quantum coherence is an essential physical resource which can be used to implement
various quantum tasks such as quantum computing \cite{Hill}, cryptography \cite{Cole}, information processing \cite{Stre1, Stre2, Diaz},
thermodynamics \cite{Lost}, metrology \cite{Fro} and quantum biology \cite{Stre3}. Various efforts have been made to build the resource theory of coherence \cite{Abe, Gour1, Lev, Bau}. The resource theory of coherence consists of two fundamental elements: free states and free operations.
Free states are diagonal quantum state in the priori fixed reference basis. Physically, we can prepare free states with no additional costs.
Free operations catch  physical transformations which can be carried out without consumption of resources. Having confirmed the two properties, people
initiate investigation of the corresponding theory like coherence manipulation and coherence quantification.
Quantitative and operational description are chief virtues of the resource theory of coherence.

Let us start by recapitulating the fundamental framework of the resource theory of quantum coherence \cite{Bau}.
Free states are defined as diagonal states in a prefixed basis  $\{|i\rangle\}_{i=1}^d$
for a $d$ dimensional Hilbert space ${\mathcal H}$, i.e., $$\rho=\sum_{i=1}^d\lambda_i|i\rangle\langle i|, $$ $\lambda_i$
is a probability distribution. The family of incoherent states will be denoted as ${\mathcal I}$. The basis $\{|i\rangle\}_{i=1}^d$
is also called incoherent basis which is chosen depending on physical problem under study \cite{Zur}. All other states which are not diagonal in
this basis are called coherent states.

Different definitions of free operations within resource  theory of quantum coherence
have been investigated  due to different physical considerations. One kind of key free operations, named incoherent operations $(\text{ICO})$, are
specified by a set of Kraus operators $\{K_j\}$ such that $K_j\rho K_j^*/Tr(K_j\rho K_j^*)\in {\mathcal
I}$ for all $\rho\in {\mathcal I}$, $$\Phi(\rho)=\sum_{j}K_j \rho K_j^*.$$ Such a definition guarantees that, even if one implements
postselection on the measurement outcomes, one can not create coherent states from an incoherent state \cite{Bau}. The Kraus operators $\{K_j\}$ are called incoherent. An incoherent operation $(\text{ICO})$ is called strictly incoherent $(\text{SIO})$ if both $K_j$ and $K_j^*$ are incoherent \cite{Win, Lud, Gour2, Gu}. Other kinds of
definitions of free operations  have been studied, like  maximally incoherent operations
$(\text{MIO})$ \cite{Abe}, physically incoherent operations
$(\text{PIO})$ \cite{Gour3}, dephasing covariant incoherent operations $(\text{DIO})$ \cite{Gour3,Mar1},   genuinely incoherent operations $(\text{GIO})$ \cite{Vice}.

 Although the resource theory of quantum coherence has been found widespread applications in different practical scenarios \cite{Stre3},
there are no physically convincing free operations selected out, setting against the central role of local operations and classical communication in the resource theory of quantum entanglement \cite{Hor4}.

 The category of strictly incoherent operations $(\text{SIO})$ has appeared to be the most potential candidate of free operations
 fulfilling desired norms of resource theory while in the meantime originated from physical motivation and experimentally enforceable
\cite{Win, Lud, Gour2, Gu, Ben, Was}. However, SIOs are specified by Kraus operators and so have various Kraus representations.
This makes execution difficult for SIOs, because the common way of executing a SIO is to add a auxiliary system, evolve the composite system
 with a unitary operator and ultimately tracing off the auxiliary system. Particularly, structural classification for unitary operators and
 auxiliary systems generating SIOs is absent. But, execution of SIOs is extremely important because it determines the whole structure of coherence manipulation.

An efficient way to comprehend the structure of SIOs is to seek out their parametrization representation.
To that end, upper bounds on the number of incoherent Kraus operators for SIOs are given \cite{Ben}.
It is shown that every single-qubit SIO has a representation with at most 4 strictly incoherent Kraus operators (and this is the optimal number).
The typical choice of the incoherent Kraus operators is provided by the family
$$\left\{\left(\begin{array}{cc}
     a_1 & 0\\
     0  & b_1\end{array}\right),\left(\begin{array}{cc}
     0 & b_2\\
     a_2  & 0\end{array}\right),\left(\begin{array}{cc}
     a_3 & 0\\
     0  & 0\end{array}\right), \left(\begin{array}{cc}
     0 & 0\\
     a_4  &0\end{array}\right) \right\},\eqno(1)
$$
where $a_i\in {\mathbb R}$,  $b_i \in {\mathbb C} $,
$\sum_{i=1}^4 a_i^2=\sum_{i=1}^2|b_i|^2=1.$


Note that Pauli operators are  incoherent,
an interesting question is to represent SIOs in terms of Pauli operators.
There are  paradigmatic instances of SIOs which are represented by Pauli operators
whose behaviour reflect classic noise sources in quantum
information processing \cite{Bau, Nie, Thomas}. The
bit-flip, bit+phase-flip, and phase-flip operations admit Kraus decomposition by

$$K_0^{F_k}=\sqrt{1-\frac{q}{2}}I,\hspace{0.1in} K_{i,j\neq k}^{F_k}=0,\hspace{0.1in} K_k^{F_k}=\sqrt{\frac{q}{2}}\sigma_k,$$
k=1,2,3, $q\in[0,1]$ and $\sigma_j$ is the
$j$th Pauli operator. The depolarizing operation ${\mathcal L}_q$
is represented in Kraus decomposition by $$\left\{\sqrt{1-\frac{3q}{4}}I,\sqrt{\frac{q}{4}}\sigma_1, \sqrt{\frac{q}{4}}\sigma_2, \sqrt{\frac{q}{4}}\sigma_3\right\}.$$

A relevant achievement in resource theory of entanglement is the parametric representation of quantum state for a two-qubit system \cite{Fano, Luo}. It is shown that every quantum state $\rho$ can be parametrized as
$$\rho =\frac{1}{4}(I+\overrightarrow{u}\overrightarrow{\sigma}\otimes I+I\otimes\overrightarrow{v}\overrightarrow{\sigma}+\sum_{i,j=1}^3 w_{ij}\sigma_i\otimes\sigma_j).$$
Here $I$ is the identity operator, $\overrightarrow{\sigma}=(\sigma_1, \sigma_2, \sigma_3)$
with $\sigma_i$ being the Pauli operators. $$\overrightarrow{u}=(u_1, u_2, u_3)\in {\mathbb R}^3, \overrightarrow{v}=(v_1, v_2, v_3)\in {\mathbb R}^3,$$ $$\overrightarrow{u}\overrightarrow{\sigma}=u_1\sigma_1+u_2\sigma_2+u_3\sigma_3,$$ etc., and $w_{ij}$ are real numbers.

It is surprising that we will find Pauli operators are not enough to represent SIOs in general. The phase operator which is  incoherent
plays an important role for parametric representation of SIOs on a single qubit.
In the following, we will re-parametrize the typical form of SIO in terms of $\{\sigma_1, \sigma_2, \sigma_3, S\}$, here $\sigma_1,\sigma_2, \sigma_3, S$ denote the Pauli operators and the phase operator respectively.
Recall that $$\sigma_1=\left(\begin{array}{cc}
                     0& 1\\
                     1& 0\end{array}\right), \sigma_2=\left(\begin{array}{cc}
                     0& -i\\
                     i& 0\end{array}\right),$$ $$\sigma_3=\left(\begin{array}{cc}
                     1& 0\\
                     0& -1\end{array}\right), S=\left(\begin{array}{cc}
                     1& 0\\
                     0& i\end{array}\right).$$

 Assume
$$\Phi\sim\left\{\left(\begin{array}{cc}
     a_1 & 0\\
     0  & b_1\end{array}\right),\left(\begin{array}{cc}
     0 & b_2\\
     a_2  & 0\end{array}\right),\left(\begin{array}{cc}
     a_3 & 0\\
     0  & 0\end{array}\right), \left(\begin{array}{cc}
     0 & 0\\
      a_4  &0\end{array}\right) \right\},$$  we only discuss that case  $\Phi$ is bistochastic for  application, i.e., $\Phi(I)=I$.

Our main result is stated as follows.

{\bf Theorem 1.} {\it Denote $a_1b_1+a_2\overline{b_2}=a+bi,(a_1b_1-a_2\overline{b_2})i=c+di,$ then                
$$\begin{array}{lll}
\Phi(\rho) &= &(\frac{1}{2}-\frac{a}{2}-\frac{b}{2}-\frac{c}{2}-\frac{d}{2}-\frac{|b_1|^2-|b_2|^2}{2})I\\
           &+ &\frac{a+d+|b_1|^2-|b_2|^2}{2}\rho+\frac{b}{2}S\rho S^*+\frac{c}{2}S^*\rho S\\
           &+&\frac{a}{2}\sigma_1\rho\sigma_1+\frac{d}{2}\sigma_2\rho\sigma_2+\frac{|b_1|^2-|b_2|^2}{2}\sigma_3\rho\sigma_3\\
           &+ &\frac{b}{2}S\sigma_1\rho\sigma_1S^*+\frac{c}{2}S^*\sigma_2\rho\sigma_2 S.\\

\end{array}$$}

 Given a quantum operation $\Phi$, the corresponding Schrodinger-picture operation ${\Phi}^*$ is defined via the duality $tr[\Phi^*(A)B] = tr[A\Phi(B)]$. It follows that $\Phi^*(\rho)=\sum_j K_j^*\rho K_j$ in terms of Kraus operators. The family $\{\Phi^n\}_{n\in N^+}$ is a discrete-time quantum-dynamical semigroup generated by $\Phi$, i.e.,
$$\Phi^ n\Phi^m =\Phi^{n+m}.$$ Physically, a quantum dynamical semigroup
describes a Markovian evolution in discrete time.

The reestablishment of quantum thermal averages by a quantum computer has been one of research focus of quantum simulation
in the past twenty years \cite{Ter, Poui, Bilg, Temm, Riera, Ge, Chow, Clem, MMotta}. In the seminal paper \cite{Ter}, the authors show how to
apply quantum computation to the research of thermodynamical properties of a single system.
 A key step for preparation of
the equilibrium state on a quantum computer is to determine whether the dynamics generated by $\Phi$ is relaxing \cite{Ter},
i.e., whether there exists a density operator $\sigma$ such that, for any density operator $\rho$, the orbit ${\Phi}^n(\rho)$ converges to $\sigma$ in the trace norm. Liapunov's theorem \cite{Streater1, Streater2, Ragin} provides a
common way to show that a dynamics is relaxing. That is, if the bistochastic channel $\Phi^*$ is ergodic $$S[\Phi^*(\rho)]-S[\rho]\geq \frac{\gamma}{2}\|\rho-\frac{1}{d}I\|_2^2,$$ $\text{where}\hspace{0.1in}\gamma\in[0,1), S(\rho)=-tr(\rho \ln\rho)$, then the dynamics generated by $\Phi$
is relaxing from Liapunov's theorem. But the method is often difficult to apply  due to the computational complexity of von Neuman entropy and spectral gap $\gamma$. Basing on Theorem 1, we can determine completely the relaxing of SIOs on a single qubit.

{\bf Theorem 2.} {\it Let $\lambda_1$ and $\lambda_2$ be the eigenvalues of
$\left(\begin{array}{cc}
             a & b \\
             c& d \\
             \end{array}\right)$, here $a, b, c, d$ are from Theorem 1, then as $n\rightarrow +\infty$, $${\Phi}^n(\rho)\rightarrow \frac{1}{2}I \Leftrightarrow|\lambda_i|<1,(i=1,2), b_1b_2\neq 0.$$}

By a direct computatin, one can see that $$\hspace{0.2in}F_1\sim\left\{\left(\begin{array}{cc}
\sqrt{1-\frac{q}{2}} & 0\\
     0  &\sqrt{1-\frac{q}{2}}\end{array}\right),\left(\begin{array}{cc}
     0 &\sqrt{ \frac{q}{2}}\\
\sqrt{\frac{q}{2}}  & 0\end{array}\right)\right\},$$

$$\hspace{0.2in}F_2\sim\left\{\left(\begin{array}{cc}
\sqrt{1-\frac{q}{2}} & 0\\
     0  &\sqrt{1-\frac{q}{2}}\end{array}\right),\left(\begin{array}{cc}
     0 &-\sqrt{ \frac{q}{2}}\\
\sqrt{\frac{q}{2}}  & 0\end{array}\right)\right\},$$

$$F_3\sim\left\{\left(\begin{array}{cc}
1-q & 0\\
     0  &1\end{array}\right),\hspace{0.1in}\left(\begin{array}{cc}
     \sqrt{2q-q^2} &0\\
0  & 0\end{array}\right)\right\}.$$
From Theorem 2, we can obtain that the bit-flip, bit+phase-flip and phase-flip operations are not relaxing. We can adjust the  coefficients of Kraus operators of $F_1$ such that
$$F_1^\theta\sim\left\{\left(\begin{array}{cc}
\sqrt{1-\frac{q}{2}} & 0\\
     0  &\sqrt{1-\frac{q}{2}}e^{i\theta}\end{array}\right),\left(\begin{array}{cc}
     0 &\sqrt{\frac{q}{2}}e^{i\theta}\\
\sqrt{\frac{q}{2}}  & 0\end{array}\right)\right\}.$$
Using Theorem 2 again, one can see $F_1^\theta$ is relaxing if and only if $|\cos\theta|\neq 1$. Furthermore, the depolarizing operation ${\mathcal L}_q$ is also relaxing. Therefore $F_1^\theta(|\cos\theta|\neq 1)$ and ${\mathcal L}_q$ can be used to preparation of equilibrium state and computation of correlation functions on a quantum computer \cite{Ter}.

\vspace{0.1in}

In Theorem 1, if $b_i\in{\mathbb R}, i=1,2$, then we can represent directly SIOs in terms of Pauli operators.

{\bf Theorem 3.} {\it  Denote $a_1b_1+a_2b_2=a, a_1b_1-a_2b_2=d,$ then
$$\begin{array}{lll}
\Phi(\rho) &= &(\frac{1}{2}-\frac{a}{2}-\frac{d}{2}-\frac{|b_1|^2-|b_2|^2}{2})I\\
           &+ &\frac{a+d+|b_1|^2-|b_2|^2}{2}\rho\\
           &+&\frac{a}{2}\sigma_1\rho\sigma_1+\frac{d}{2}\sigma_2\rho\sigma_2+\frac{|b_1|^2-|b_2|^2}{2}\sigma_3\rho\sigma_3.\\
          \end{array}$$}

The coherence manipulation is basic in the resource theory of coherence.
Given two coherent states $\rho$ and $\sigma$, the question of coherence manipulation is to study whether there exists some free operation $\Phi$
such that $\Phi(\rho)=\sigma$.
Attempts have been made to uncover the fundamental laws of its behaviour under free operations \cite{ Stre3, Bau, Du1, Win, Gour2, Ben,  Ma, Du2, Lzhou}.  In \cite{Ben}, it is proved that, for qubit states $\rho, \sigma$ with Bloch vector
$r=(r_x,r_y,r_z)^t$ and $s=(s_x,s_y,s_z)^t$,
$\rho$ can be converted into $\sigma$ by stochastic SIO if and only if the following inequalities are fulfilled $${s_x}^2+{s_y}^2\leq {r_x}^2+{r_y}^2, \hspace{0.1in}|s_z|\leq |r_z|.$$ From the proof of Theorem 1, one can see that for qubit states $\rho, \sigma$ with Bloch vector
$r=(r_x,r_y,r_z)^t$ and $s=(s_x,s_y,s_z)^t$,
$\rho$ can be transformed into $\sigma$ by stochastic SIO with the same decomposition as Theorem 3 if and only if  $$|s_x|\leq |r_x|, \hspace{0.1in} |s_y|\leq |r_y|,\hspace{0.1in}|s_z|\leq |r_z|.$$ This implies that the operational ability of stochastic SIO represented by Pauli operators and the general stochastic SIO is different. An efficient method in coherence manipulation is to characterize the image of some coherent state under all some
kind of  free operations \cite{Ben}. Therefore the image of $\rho$ under all stochastic SIOs as Theorem 1 is a cylinder while the image of $\rho$ under all stochastic SIOs as Theorem 3 is a cuboid.


\vspace{0.1in}
{\it Proofs of results.---}
{\bf Proof of Theorem 1.}  Let ${\mathcal M}_2$ be the set of $2\times 2$ matrices and $\hspace{0.1in}{\mathcal X}=\{A: A\in{\mathcal M}_2, A=A^*, tr(A)=0\}.$
We will regard ${\mathcal X}$ as a real vector space which
is in fact a Hilbert space endowed with the
inner product $\langle A, B\rangle= tr(AB)$.  In fact, ${\mathcal X}$ is a real Hilbert subspace  consisting
of all self-adjoint operators which are orthogonal to $I$. Since $\Phi$ is trace preserving,
$\Phi$ can be regarded as a linear operator on ${\mathcal X}$ and has a matrix representation with respect to the orthogonal basis $\sigma_1, \sigma_2, \sigma_3$ of ${\mathcal X}$.
From $\Phi(I)=I$, we have $$a_1^2+a_3^2+|b_2|^2=1\hspace{0.1in}{\text and}\hspace{0.1in} a_2^2+a_4^2+|b_1|^2=1.$$
By a direct computation, one can obtain
$$\Phi=\left(\begin{array}{ccc}
             a & c & 0\\
             b & d & 0\\
             0 & 0 & |b_1|^2-|b_2|^2\end{array}\right)$$ under the basis vectors $\sigma_1, \sigma_2, \sigma_3$.
Let $$\Phi_{11}(\cdot)=\frac{1}{2}\cdot+\frac{1}{2}\sigma_1\cdot\sigma_1.$$ It is easy to see that
$$\Phi_{11}(\sigma_1)=\sigma_1 \hspace{0.1in}\text{and} \hspace{0.1in}\Phi_{11}(\sigma_2)=\Phi_{11}(\sigma_3)=0.$$
Similarly,  let $$\Phi_{22}(\cdot)=\frac{1}{2}\cdot+\frac{1}{2}\sigma_2\cdot\sigma_2$$ and $$\Phi_{33}(\cdot)=\frac{1}{2}\cdot+\frac{1}{2}\sigma_3\cdot\sigma_3.$$ One can check that
$$\Phi_{22}(\sigma_2)=\sigma_2,\ \  \Phi_{22}(\sigma_1)=\Phi_{22}(\sigma_3)=0,$$
$$\Phi_{33}(\sigma_3)=\sigma_3,\ \ \Phi_{33}(\sigma_1)=\Phi_{33}(\sigma_2)=0.$$
Let $$\Phi_{21}(\cdot)=S\Phi_{11}(\cdot)S^*.$$ From $S\sigma_1 S^*=\sigma_2$, we have
$$\Phi_{21}(\sigma_1)=\sigma_2, \ \ \Phi_{21}(\sigma_2)=\Phi_{21}(\sigma_3)=0.$$
Analogously, set $$\Phi_{12}(\cdot)=S^*\Phi_{22}(\cdot)S.$$ One have $$\Phi_{12}(\sigma_2)=\sigma_1,\ \
\Phi_{12}(\sigma_1)=\Phi_{12}(\sigma_3)=0.$$ Therefore $$\Phi=a\Phi_{11}+b\Phi_{21}+c\Phi_{12}+d\Phi_{22}+(|b_1|^2-|b_2|^2)\Phi_{33}.$$
By a direct computation,  one can obtain $$\begin{array}{lll}
       \Phi(\rho)&=&\Phi(\rho-\frac{1}{2}I)+\Phi(\frac{1}{2}I)\\
                 &=&a\Phi_{11}(\rho-\frac{1}{2}I)+b\Phi_{21}(\rho-\frac{1}{2}I)\\
                 & &+c\Phi_{12}(\rho-\frac{1}{2}I)+d\Phi_{22}(\rho-\frac{1}{2}I)\\
                 & &+(|b_1|^2-|b_2|^2)\Phi_{33}(\rho-\frac{1}{2}I)+\frac{1}{2}I\\
                 &=&(\frac{1}{2}-\frac{a}{2}-\frac{b}{2}-\frac{c}{2}-\frac{d}{2}-\frac{|b_1|^2-|b_2|^2}{2})I\\
                 & &+\frac{a+d+|b_1|^2-|b_2|^2}{2}\rho+\frac{b}{2}S\rho S^*+\frac{c}{2}S^*\rho S\\
                 & &+\frac{a}{2}\sigma_1\rho\sigma_1+\frac{d}{2}\sigma_2\rho\sigma_2+\frac{|b_1|^2-|b_2|^2}{2}\sigma_3\rho\sigma_3\\
                 & &+\frac{b}{2}S\sigma_1\rho\sigma_1S^*+\frac{c}{2}S^*\sigma_2\rho\sigma_2 S.
                 \end{array}$$

{\bf Proof of Theorem 2.} From the proof of Theoren 1, we have $$\Phi=\left(\begin{array}{cccc}
             a & c & 0& 0\\
             b & d & 0& 0\\
             0 & 0 & |b_1|^2-|b_2|^2& 0\\
              0 & 0 & 0&1
             \end{array}\right)$$ under the orthogonal basis  $\sigma_1, \sigma_2, \sigma_3$, $I$ of ${\mathcal M}_2$.
In \cite[Theorem 1]{Tanzhi}, the general calculation formula  of the $n-th$ power of $2\times 2$ matrix is proposed. With the help of the formula,
we can compute $\Phi^n$ which is treated in three cases.

Case 1. $(a-d)^2+4bc> 0$.

It is easy to see that $\lambda_1\neq \lambda_2$ and $\lambda_i\in{\mathbb R}, i=1,2$. Furthermore
$$\begin{array}{ll}
&\Phi^n \\
=& \left(\begin{array}{cc}
                      \frac{\lambda_{1}^{n+1}-\lambda_{2}^{n+1}}{\lambda_1-\lambda_2}-
                      d\frac{\lambda_{1}^{n}-\lambda_{2}^{n}}{\lambda_1-
                      \lambda_2} & c\frac{\lambda_{1}^{n}-\lambda_{2}^{n}}{\lambda_1-\lambda_2}\\
                      b\frac{\lambda_{1}^{n}-\lambda_{2}^{n}}{\lambda_1-\lambda_2}&\frac{\lambda_{1}^{n+1}-\lambda_{2}^{n+1}}{\lambda_1-\lambda_2}-
                      a\frac{\lambda_{1}^{n}-\lambda_{2}^{n}}{\lambda_1-
                      \lambda_2}\end{array}\right)\end{array}$$
$$\oplus \left(\begin{array}{cc}
(|b_1|^2-|b_2|^2)^n& 0\\
0 & 1\end{array}\right).$$
One can see that $\lim\limits_{n\rightarrow +\infty}{\Phi}^n(\rho)=\frac{1}{2}I$  for any density operator $\rho$ if and only if
the limit of each entry of ${\Phi}^n$ is zero except $(4,4)$ position. We can deduce that $b_1b_2\neq 0$ since $|b_1|^2+|b_2|^2=1$. By \cite[Proposition 1]{Ter}, we know $|\lambda_i|\leq 1, i=1,2$. If $b=0, c=0$,
then ${\Phi}^n$ is diagonal, and so $|\lambda_i|<1$. If $b\neq 0$ or $c\neq 0$, then $\lim\limits_{n\rightarrow +\infty} ({\lambda_1}^{n}-{\lambda_2}^{n})=0$. This implies $|\lambda_i|<1$.

Case 2. $(a-d)^2+4bc= 0$.

In this case, we have $\lambda_1=\lambda_2\in{\mathbb R}$.
$$\begin{array}{ll}
&{\Phi}^n \\
=& \left(\begin{array}{cc}
                      (n+1)\lambda_1^n-dn\lambda_1^{n-1} & bn\lambda_1^{n-1}\\
                      cn\lambda_1^{n-1}&(n+1)\lambda_1^n-an\lambda_1^{n-1}\end{array}\right)\end{array}$$
$$\oplus \left(\begin{array}{cc}
(|b_1|^2-|b_2|^2)^n& 0\\
0 & 1\end{array}\right).$$
 Note that $\lim\limits_{n\rightarrow +\infty}{\Phi}^n(\rho)=\frac{1}{2}I$  for any density operator $\rho$ if and only if the limit of each entry of ${\Phi}^n$ is zero except $(4,4)$ position. Thus $b_1b_2\neq 0$. From \cite[Proposition 1]{Ter}, $|\lambda_1|\leq1$ and so $|\lambda_1|<1$.

Case 3. $(a-d)^2+4bc< 0$.

It is evident $\lambda_1=\overline{\lambda_2}$. Furthermore
$$\begin{array}{ll}
&{\Phi}^n \\
=& \left(\begin{array}{cc}
                       a_{11}& b(\sqrt{ ad-bc})^{n-1}\frac{sin(n\theta)}{sin\theta}\\
                      c(\sqrt{ ad-bc})^{n-1}\frac{sin(n\theta)}{sin\theta}&a_{22}\end{array}\right)\end{array}$$
$$\oplus \left(\begin{array}{cc}
(|b_1|^2-|b_2|^2)^n& 0\\
0 & 1\end{array}\right),$$
$$a_{11}=(\sqrt{ ad-bc})^n\frac{sin(n+1)\theta}{sin\theta}-d\sqrt{ ad-bc})^{n-1}\frac{sin(n\theta)}{sin\theta},$$
$$a_{22}=(\sqrt{ ad-bc})^n\frac{sin(n+1)\theta}{sin\theta}-a\sqrt{ ad-bc})^{n-1}\frac{sin(n\theta)}{sin\theta}.$$
By similar argument as case 1 and case 2, we have $b_1b_2\neq 0$ and $|ad-bc|<1$.
Note that $\lambda_1\lambda_2=|\lambda_1|^2=|\lambda_2|^2=ad-bc$. Thus $|\lambda_i|<1$, as desired.

\vspace{0.1in}
{\it Conclusion.---}
The efficient description of SIOs on a single qubit is offered in physical dimensional. In entanglement theory, it is wellknown that the family of local operation and classical communication is notoriously difficult to catch mathematically \cite{Leung}. We also apply our results to
reconstruction of quantum thermal averages and coherence manipulation. Coherence manipulation in the qubit case of is especially important since it can be demonstrated experimentally \cite{Guo1}.

\vspace{0.1in}
{\it Acknowledgement.---}
We acknowledge that the research was  supported by NSF of China (11671332), NSF of
Fujian (2018J01006).

\end{document}